\documentclass[aps, superscriptaddress, floatfix]{revtex4}

\usepackage{graphicx}

\begin{document}



\newcommand{\Dirac}{\rlap{\hspace{-.5mm} \slash} D}
\newcommand{\sumint}{\rlap{\hspace{-.5mm} $\sum$} \int}
\title{
 QCD Phase Transitions in the $1/N_c$ Expansion}
\author{D.~Toublan}
\affiliation {Physics Department, University of Maryland, College
Park, MD 20742}

\date{\today}

\begin{abstract}
We study the QCD phase diagram at nonzero baryon and isospin
chemical potentials using the $1/N_c$ expansion.  We find that there
are two phase transitions between the hadronic phase and the quark
gluon plasma phase.  We discuss the consequences of this result for
the universality class of the critical endpoint at nonzero baryon
and zero isospin chemical potential.
\end{abstract}

\maketitle

\section{Introduction}
In order to understand neutron stars, the early Universe, and heavy
ion collision experiments, it is necessary to better grasp the
physics of strong interacting matter in extreme conditions.
Therefore the study of the QCD phase diagram at nonzero temperature
and densities is very important. The nonperturbative lattice
simulations that successfully addressed problems at zero densities
can be used at nonzero isospin density \cite{KSisospin}, but not at
nonzero baryon density because of  the so-called "sign problem". One
has therefore to rely on novel approaches to study QCD at nonzero
baryon density \cite{lattMuB_F&K, lattMuB_Bielefeld, lattMuB_ZH,
lattMuB_Maria}, which corresponds to the most important physical
situations. In particular, these new methods have been used to study
the critical temperature that separates the hadronic phase from the
quark gluon plasma phase. However, these studies are valid only at
small chemical potentials.

In a previous article, we have successfully used the $1/N_c$
expansion of QCD to explain some key properties of the critical
temperature that separates the hadronic phase from the quark gluon
plasma phase \cite{critTempNc}. In the present work we shall extend
our study to the general case $\mu_u\neq\mu_d$ and investigate the
chiral phase transitions.  Several models have shown that the phase
diagram might be qualitatively altered in this case
\cite{qcdMuBMuI_RMT, qcdMuBMuI_NJL, qcdMuBMuI_Ladder}. We shall also
describe the consequences of our results for the universality class
of the critical endpoint. We shall restrict ourselves to phase
transitions between the hadronic phase and the quark gluon plasma
phase, i.e. to situations with $\mu_I< m_\pi$, avoiding conditions
where the ground state becomes a pion superfluid at $T=0$ \cite{KST,
SS, KT, STV}.

\section{Critical Temperatures}
We use the usual $1/N_c$ expansion with 't~Hooft's coupling
\cite{tHooft}. As explained in detail in
\cite{cohenLargeNmuT,critTempNc}, the $1/N_c$ expansion of the
pressure reads:
\begin{eqnarray}
p(T,\{m_f,\mu_f\})=N_c^2 \Big(p_0(T)+ \frac1{N_c} \sum_{f=1}^{N_f}
p_1(T,m_f,\mu_f^2) + {\cal O}(\frac1{N_c^2}) \Big),
\end{eqnarray}
where $f$ are the different light quark flavors.  The sets of
diagrams that contribute to the pressure are shown in Fig.~1.
\begin{figure}[h]
\hspace{-.8cm}
\includegraphics[scale=0.75, clip=true, angle=0,
draft=false]{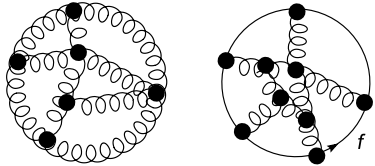} \caption{\label{pressure} Sets of
diagrams that contribute to the pressure.  The symbol $f$ denote the
quark flavor and runs over all flavors. The first diagram is ${\cal
O}(N_c^2)$, and the second is ${\cal O}(N_c)$.}
\end{figure}
In a finite volume, the specific heat will peak at the transition
between the hadronic phase and the quark gluon plasma phase.  This
peak might diverge in the thermodynamic limit, depending on whether
there is a genuine phase transition or only a crossover.  The
specific heat can be derived directly from the pressure leading to
\cite{critTempNc}
\begin{eqnarray}
\label{cv} C_V=N_c^2 \Big(c_0(T)+ \frac1{N_c} \sum_{f=1}^{N_f}
c_1(T,m_f,\mu_f^2) + {\cal O}(\frac1{N_c^2}) \Big).
\end{eqnarray}
Therefore, the critical temperature, which can be obtained by
solving $\partial C_V/\partial T|_{T_c}=0$ is given by
\cite{critTempNc}:
\begin{eqnarray}
\label{tc} t_c=\frac1{N_c} \sum_{f=1}^{N_f} f(\mu_f^2) +{\cal
O}(\frac1{N_c^2}),
\end{eqnarray}
where the reduced temperature $t_c=(T_c-T_0)/T_0$, with $T_0$ the
critical temperature at zero chemical potentials.

The chiral susceptibilities, $\chi_f=\partial^2 p/\partial m_f^2$,
peak at the transition or crossover where chiral symmetry is
partially restored (chiral symmetry cannot be completely restored
since we restrict ourselves to the case $m_f\neq0$).  The diagrams
that contribute to $\chi_f$ up to next-to-next-to-leading order are
shown in Fig.~2.
\begin{figure}[h]
\hspace{-.8cm}
\includegraphics[scale=0.75, clip=true, angle=0,
draft=false]{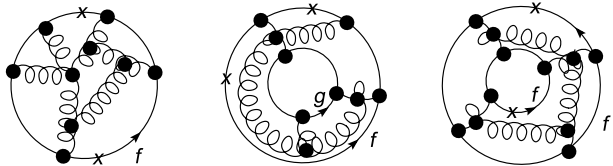} \caption{\label{susc} Sets of diagrams that
contribute to the chiral susceptibility $\chi_f$.  The crosses are
insertions of the mass operator, $f$ is the specific quark flavor
associated with $\chi_f$, and $g$ runs over all quark flavors. The
first diagram is ${\cal O}(N_c)$, the second and third are ${\cal
O}(N_c^0)$.}
\end{figure}
The $1/N_c$ expansion of the chiral susceptibilities can be
expressed as
\begin{eqnarray}
\label{chi} \chi_f&=&N_c \Big( \chi_0(T,\mu_f^2) +\frac1{N_c} \Big(
N_f \chi_1(T,\mu_f^2) +  \sum_{g=1}^{N_f} \big(\chi_2(\mu_g^2)
 + \chi_3(\mu_f \mu_g) \big) + \chi_4(\mu_f^2) \Big) +{\cal O}(\frac1{N_c^2}) \Big),
\end{eqnarray}
where $\chi_0$ comes from the first diagram in Fig.~2,
$\chi_{1,2,3}$ come from the second diagram in Fig.~2 with the
$\mu$-dependence taken respectively from the outside quark loop only
($\chi_1$), the inside quark loop only ($\chi_2$), and both the
inside and outside quark loops ($\chi_3$), and $\chi_4$ comes from
the third diagram in Fig.~2. In the equation above, we have used
that $p(\mu_u,\mu_d)=p(-\mu_u,-\mu_d)$ because of CP, and thus that
$\chi_f(\mu_u,\mu_d)=\chi_f(-\mu_u,-\mu_d)$, and that for equal
masses, $\chi_u(\mu_u,\mu_d)=\chi_d(\mu_d,\mu_u)$. If we restrict
ourselves to $m_u=m_d$, notice that the that these latter two
properties imply that $\chi_u=\chi_d$ for $\mu_u=\pm\mu_d$, but that
$\chi_u\neq\chi_d$ in general since $\mu_u^2\neq\mu_d^2$. Therefore,
since the critical temperature for the restoration of chiral
symmetry is defined by a peak in $\chi_f$, each flavor might have a
different critical temperature when $\mu_u^2\neq\mu_d^2$.

The critical temperature that corresponds to the restoration of
chiral symmetry, $T_f$, can be obtained from
$\partial\chi_f/\partial T|_{T_f}=0$. Therefore, we find that the
reduced temperature for chiral symmetry restoration,
$t_f=(T_f-T_0)/T_0$, is given by
\begin{eqnarray}
\label{tf}
 t_f&=&\tau_0(\mu_f^2)+\frac1{N_c}
 \Big( N_f \tau_1(\mu_f^2) + \sum_{g=1}^{N_f} \big(\tau_2(\mu_g^2)
 + \tau_3(\mu_f \mu_g) \big) +\tau_4(\mu_f^2) \Big)+{\cal O}(\frac1{N_c^2}).
\end{eqnarray}

Lattice simulations have shown that there is only one phase
transition when $\mu_u=\pm\mu_d$, at least at small chemical
potentials for $N_c=2$ and $3$ \cite{lattMuB_Bielefeld, lattMuB_ZH,
KSisospin, lattNc2}. We assume that this property holds at large
$N_c$. Therefore the critical temperatures $t_c$ and $t_f$ for each
flavor should coincide, at least at small chemical potential.
However, it is clear from (\ref{tc}) and (\ref{tf}) that $t_c$ and
$t_f$ do not have the same $1/N_c$ expansion: the chemical potential
enters at a different order in the $1/N_c$ expansion, and, at
leading order, $t_c$ depends equally on all chemical potentials,
whereas $t_f$ depends only on one chemical potential. As was shown
in \cite{critTempNc}, the qualitative properties of $t_c$ are in
agreement with the lattice results. We therefore impose that
$t_c=t_u=t_d$ for $\mu_u=\pm\mu_d$ as a function of $N_f$ and $N_c$
in the $1/N_c$ expansion. These constraints imply that
\begin{eqnarray}
\label{constraint}
\tau_0(\mu^2)&=&0 \nonumber \\
\tau_3(\mu^2)&=&\tau_3(-\mu^2) \\
\tau_4(\mu^2)&=&0 \nonumber \\
f(\mu^2)&=&\tau_1(\mu^2)+\tau_2(\mu^2)+\tau_3(\mu^2). \nonumber
\end{eqnarray}
Notice that the same conclusions for QCD with an even number of
flavors can be reached by using the dependence of $t_c$ and $t_f$ on
the quark masses rather than on the chemical potentials.

In general, we expect that $\tau_{1,3}\neq0$.  We therefore conclude
that critical temperatures for two different flavors will be related
by
\begin{eqnarray}
\label{tempDiff}
 t_f-t_g=\frac1{N_c} \Big( N_f(\tau_1(\mu_f^2)-\tau_1(\mu_g^2))
 + \sum_{h=1}^{N_f} ( \tau_3(\mu_f\mu_h) - \tau_3(\mu_g\mu_h) )\Big)
 +{\cal O}(\frac1{N_c^2}),
\end{eqnarray}
and that the critical temperatures are sensitive to {\it all}
chemical potentials at the {\it same} order in the $1/N_c$
expansion. Therefore the restoration of the chiral symmetry takes
place at different temperatures at $\mu_u^2\neq\mu_d^2$, i.e. at
nonzero baryon {\it and} isospin chemical potentials.  This is
important since most experiments are precisely done in these
conditions.  A similar relation between the critical temperatures
for different flavors can also been written as a function of the
quark masses at zero chemical potentials.

\section{Phase Diagram}
We shall now analyze the consequences of this $1/N_c$ analysis for
the QCD phase diagram.  We shall restrict ourselves to small
positive chemical potentials, i.e. $\mu_u$, $\mu_d>0$. Possible
generic phase diagrams resulting from the above analysis are
sketched in Fig.~3.
\begin{figure}[h]
\hspace{-.8cm}
\includegraphics[scale=0.66, clip=true, angle=0,draft=false]{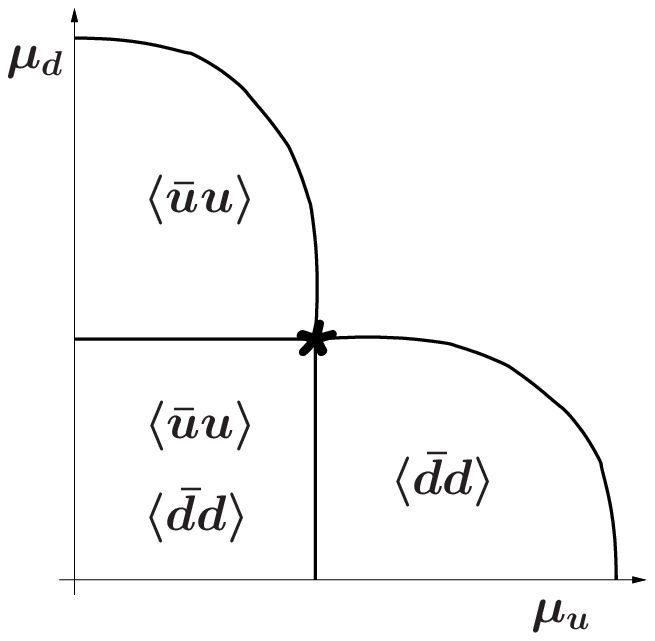}
\hspace{1.5cm}\includegraphics[scale=0.66, clip=true,
angle=0,draft=false]{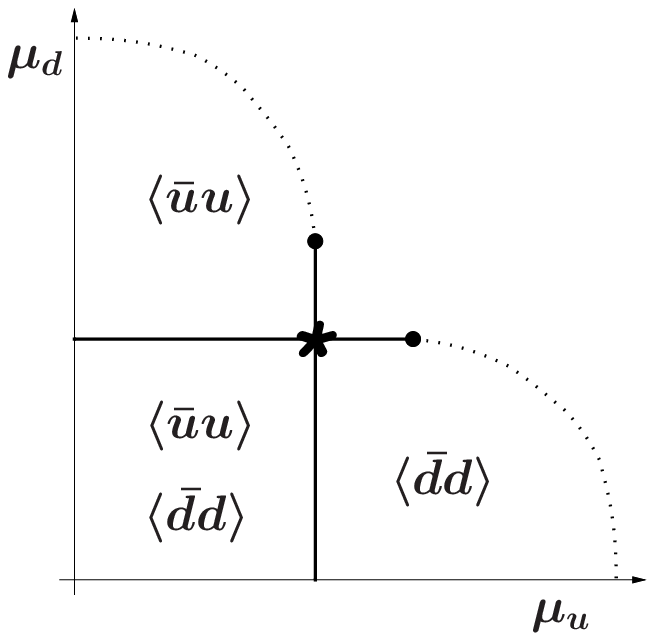} \caption{\label{phaseD}
Generic phase diagrams at fixed $T$ in the $(\mu_u,\mu_d)$ plane in
the $1/N_c$ expansion. The chiral condensates are indicated where
they are large.  The solid curves are first order phase transitions,
and the dotted curves are crossovers. The dots are critical
endpoints, and the stars denote the points that become tetracritical
endpoints at some temperature. See text for more details.}
\end{figure}

In the phase diagrams presented in Fig.~3, the temperature is high
enough so that the pion condensation phase is absent for $\mu_u$,
$\mu_d>0$ (or occupies at most a small portion of the phase diagrams
near the $\mu_u$ and $\mu_d$ axes that we will ignore here). The
temperature in these diagrams is below the temperature of the
critical endpoint at $\mu_u=\mu_d$, $T_e$  \cite{ladder, NJL, RMT}.
Above this temperature, all the transitions become crossovers.  The
points denoted by stars in the phase diagrams of Fig.~3 become
tetracritical endpoints at $T=T_e$.  The universality class of these
points is that of the $Z_2\times Z_2$ Ising model, since there are
two massless sigma modes that correspond to the divergence of both
the baryon and isospin susceptibilities.

It is possible that accidental circumstances might alter the generic
phase diagrams presented in Fig.~2.  If $\tau_1=0$ and $\tau_3=0$,
the chiral phase transitions would take place at the same
temperature and chemical potentials for all flavors. Another
accident could lead to the merging of the two critical endpoints in
the second phase diagram of Fig.~3 into one single point at the
corner of the region where all chiral condensates are large. In this
case, the critical endpoint at $T_e$ would be in the universality
class of the $Z_2$ Ising model, with only one massless sigma mode.
However, we consider these latter two phase diagrams to be
accidental in the $1/N_c$ expansion: They are accidents in the same
way as $\tau_0=0$ is an accident.

Similar phase diagrams as those presented in Fig.~3 can also be
obtained from a slightly modified version of a Random Matrix model
\cite{RMT, qcdMuBMuI_RMT}. In the original model, at sufficiently
high temperatures, the pion condensation phase never appears for
$\mu_u$, $\mu_d>0$ \cite{qcdMuBMuI_RMT, qcdMuBMuI_NJL,
qcdMuBMuI_Ladder} and the flavor sectors decouple from each other.
We use the Random Matrix effective potential and artificially add
flavor-mixing terms to it:
\begin{eqnarray}
\Omega_{RMT}&=&(\sigma_u-m)^2+(\sigma_d-m)^2 \\
&&\hspace{-1cm}-\frac14 \log \Big[ \big(((\sigma_u^2-(\mu_u^2+a
\mu_d^2)+T^2)^2+4 T^2 \mu_u^2) ((\sigma_d^2-(\mu_d^2+a
\mu_u^2)+T^2)^2+4 T^2 \mu_d^2) +b (\mu_u^2-\mu_d^2)^2\sigma_u^2
\sigma_d^2 \big)^2
 \Big], \nonumber
\end{eqnarray}
where $\sigma_{u,d}$ represent the chiral condensates.  Depending on
the choice of the artificially introduced parameters $a$ and $b$,
one can obtain either of the phase diagrams presented in Fig.~3.

\section{Conclusions}
In this article we have used the $1/N_c$ expansion to study the QCD
phase diagram at nonzero baryon and isospin chemical potentials,
which corresponds to the most common experimental conditions. We
have limited our study to the phase transitions from the hadronic
phase to the quark-gluon plasma phase.  We have also used
constraints derived from lattice simulations, in particular that the
chiral and deconfinement phase transitions take place at the same
temperature for small baryon and isospin chemical potentials.  For
QCD with two flavors, we have found that there are two chiral phase
transitions in general.  The QCD phase diagram at nonzero baryon
chemical potential is thus qualitatively different at zero and
nonzero isospin chemical potential, as other models have predicted
\cite{qcdMuBMuI_RMT, qcdMuBMuI_NJL, qcdMuBMuI_Ladder}.  This is
important since most experiments are done at nonzero baryon and
isospin chemical potentials.  As a consequence, the universality
class of the critical endpoint at nonzero baryon and zero isospin
chemical potentials should be that of the $Z_2\times Z_2$ Ising
model. The nature of the critical endpoint has observable
consequences for heavy ion collision experiments \cite{SRS}.  We
conjecture that the real phase diagram resembles the second one in
Fig.~3, because, following the analysis of Pisarski and Wilczek
\cite{pisarskiWilczek}, the restoration of chiral symmetry for one
flavor should be a crossover. The question of the deconfinement
phase transition remains open. However, we conjecture that it will
take place where the first chiral transition happens, since it is
widely believed that the deconfinement phase transition cannot
happen after a chiral phase transition.

Finally, we want to comment on the possibility to test the
predictions made above.  First the properties of the sets of
diagrams that enter into the chiral susceptibility and in particular
the relations (\ref{constraint}) can be tested on the large-$N_c$
lattice approach outlined in \cite{neubergerN}. We stress that these
tests can also be carried at zero chemical potential by studying
these diagrams as a function of the quark masses. Second, the
current lattice techniques used for $\mu_u=\mu_d$ can easily be
extended to the more general case $\mu_u\neq\mu_d$. It might however
be difficult to see the two chiral phase transitions on the lattice
in this case, since these studies are limited to small chemical
potentials and that the separation between the critical temperatures
should be proportional to $\mu_B\mu_I/N_c$.  Third, the $1/N_c$
expansion also leads to different critical temperatures for the
different flavors if they have different masses.  Recent lattice
results at zero chemical potentials have shown that the critical
temperatures indeed differ for two light quarks and one heavier
quark \cite{MILCthermo}, in agreement with the $1/N_c$ expansion
analysis presented here.

\begin{acknowledgments}
It is a pleasure to thank T.~Cohen for useful discussions.  The
Particle Physics Group at the Rensselaer Polytechnic Institute,
where part of this work was completed, is also thanked for its warm
hospitality. Work supported by the NSF under grant No.
NSF-PHY0304252.
\end{acknowledgments}

\end{document}